\documentclass[a4paper,preprintnumbers,floatfix,superscriptaddress,pra,twocolumn,notitlepage,aps]{revtex4-2}

\usepackage[utf8]{inputenc}
\usepackage[T1]{fontenc}
\usepackage{amsmath, amsthm, amssymb,amsfonts,mathbbol,amstext}
\usepackage{graphicx}
\usepackage{dcolumn}
\usepackage{bm}
\usepackage{bbm}
\usepackage{mathtools}
\usepackage{comment}
\usepackage{multirow}
\usepackage{mathrsfs}
\usepackage{physics}
\usepackage{soul}
\usepackage{adjustbox}
\usepackage{dsfont}
\usepackage[table, svgnames, dvipsnames]{xcolor}
\usepackage[colorlinks=true,citecolor=NavyBlue,linkcolor=RubineRed,urlcolor=NavyBlue]{hyperref}

\renewcommand{\braket}[3]{\langle #1 \vert #2 \vert #3 \rangle}
\providecommand{\overlap}[2]{\langle #1 \vert #2 \rangle}
\renewcommand{\ket}[1]{\vert #1 \rangle}
\renewcommand{\bra}[1]{\langle #1 \vert}

\begin{document}

\title{Krylov Complexity and Dynamical Phase Transition in the quenched LMG model}

\author{Pedro H. S. Bento} 
\affiliation{QPequi Group, Institute of Physics, Federal University of Goi\'{a}s, 74.690-900, Goi\^{a}nia, Brazil}

\author{Adolfo del Campo}
\affiliation{Department  of  Physics  and  Materials  Science,  University  of  Luxembourg,  L-1511  Luxembourg, Luxembourg}
\affiliation{Donostia International Physics Center,  E-20018 San Sebasti\'an, Spain}

\author{Lucas C. Céleri}
\affiliation{QPequi Group, Institute of Physics, Federal University of Goi\'{a}s, 74.690-900, Goi\^{a}nia, Brazil}

\begin{abstract}
Investigating the time evolution of complexity in quantum systems entails evaluating the spreading of the system's state across a defined basis in its corresponding Hilbert space. Recently, the Krylov basis has been identified as the one that minimizes this spreading. In this study, we develop a numerical exploration of the Krylov complexity in quantum states following a quench in the Lipkin-Meshkov-Glick model. Our results reveal that the long-term averaged Krylov complexity acts as an order parameter for this model. It effectively discriminates between the two dynamic phases induced by the quench, sharing a critical point with the conventional order parameter. Additionally, we examine the inverse participation ratio and the Shannon entropy in both the Krylov basis and the energy basis. A matching dynamic behavior is observed in both bases when the initial state possesses a specific symmetry. This behavior is analytically explained by establishing the equivalence between the Krylov basis and the pre-quench energy eigenbasis.
\end{abstract}

\maketitle

\section{Introduction}

Quantifying the complexity of physical phenomena has been a long-standing area of research~\cite{Zurek1990}. Various measures, such as the entropy generated during the time evolution of a quantum state~\cite{Pagels1988}, algorithmic randomness~\cite{Zurek1989}, and quantum Kolmogorov complexity~\cite{Laplante2001}, have been utilized to characterize complexity. While existing measures capture specific features, some fail to encompass other aspects and exhibit ambiguities due to the specific basis choices involved in their definitions.

Chaotic systems are expected to be more complex than integrable ones, a notion often attributed to phenomena like the butterfly effect
~\cite{Lorenz1993,Stanford2016}. In the realm of quantum systems, this effect is typically characterized by the exponential growth of the noncommutativity of local operators over time, as quantified by out-of-time order correlators. However, this exponential increase is not consistently observed in chaotic systems and may even manifest in nonchaotic ones~\cite{Fine2014,Galitski2020,Watanabe2020}.

A novel measure of complexity has recently been proposed, relying on the spread of operators across a specifically ordered basis known as the Krylov basis~\cite{Altman2019}. This basis provides the description of the dynamics in terms of a single particle moving in a semi-infinite chain, with hopping rates set by the Lanczos coefficients obtained during the computation of the Krylov basis~\cite{Lanczos1950}. A large body of literature has followed, addressing fundamental properties of this measure \cite{Barbon19,Dymarsky20,Rabinovici21,Smolkin2021,Rabinovici22,Patramanis2022,Sonner2022,delCampo2022,avdoshkin2022krylov,Muck2022,Watanabe2023, KuntalPal2023}, its generalizations for other kinds of evolution \cite{Bhattacharya2022,Zhai2023,Hornedal2023,Bhattacharjee2023,bhattacharya2023krylov,bhattacharjee2023operator}, and applications, e.g., to characterize phases of matter \cite{Trigueros2022,Caputa22,afrasiar2023}, and for quantum control \cite{takahashi2023shortcuts,bhattacharjee2023lanczos}.  
The Krylov complexity involves a choice of inner product between operators, raising questions about the optimal inner product for minimizing complexity. This challenge was addressed with the extension of the Krylov complexity for quantum state evolution presented in Ref.~\cite{Wu2022}. It was shown that the Krylov basis minimizes the proposed complexity definition. As a result,  the Krylov complexity of quantum states, also known as the spread complexity, emerges as an unambiguous measure, eliminating the need for an inner product choice. It has been explored in the SYK model~\cite{Altman2019,Rabinovici21}, the XXZ model \cite{Rabinovici22}, the Su-Schrieffer-Heeger model \cite{Caputa22}, evolutions governed by dynamical symmetry groups associated with specific Lie algebras~\cite{Patramanis2022,delCampo2022,Hornedal2023}, and random matrix models~\cite{delCampo2022,Wu2022}, among others. Additionally, it has been applied to analyze the transition from integrability to a chaotic regime~\cite{Watanabe2023, Wisniacki2023}.

The exploration of dynamics following a global quench in quantum many-body systems is widely used to probe a system in nonequilibrium statistical mechanics \cite{Eisert2015}. Global quenches induce rapid departures from equilibrium, along with coherence in the energy spectrum, producing intriguing effects, such as dynamical quantum phase transitions (DQPTs)~\cite{HeylReview,Rey2022}. These transitions arise in quenched many-body systems and manifest as cusps in the Loschmidt echo or survival probability, i.e., non-analytic behavior in the time domain for certain initial states. 
This paper contributes to their study by investigating the evolution of Krylov complexity in quantum states following a quench in the paradigmatic Lipkin-Meshkov-Glick (LMG) model~\cite{LMG1965}. Our study of the LMG model, which may exhibit a  DQPT  depending on the quench~\cite{Zunkovic2016}, reveals that the long-time averaged Krylov complexity serves as an order parameter for the DQPT. The Krylov complexity manifests distinct behavior over time in both dynamical phases induced by the quench and remarkably mirrors the oscillations and revivals seen in one of the components of the total magnetization, a conventional dynamical order parameter in spin systems.

In order to understand this fact, we conduct a numerical study of the inverse participation ratio (IPR) and the Shannon entropy over time in both the Krylov basis and the pre-quench energy basis. Remarkably, these two quantities can display identical behavior in both bases. Finally, by deriving the Krylov basis for the case of a quench starting at a null magnetic field, we analytically demonstrate its equivalence to the pre-quench energy basis. This analytical result explains the identical temporal behavior of IPR and Shannon entropy in these two bases and also why Krylov complexity is an order parameter for the considered model at null magnetic field. 

Two early studies of the Krylov complexity in the LMG model have been reported. The first one \cite{Bhattacharjee2022} addresses scrambling dominated by classical saddle points in the model, and the second \cite{afrasiar2023} identifies the spread complexity as a probe for the equilibrium quantum phase transition. However, none of these two works explores the phenomenon of DPQTs.

The paper is structured as follows: Section~\ref{sec:dynamical} provides a review of key concepts related to dynamical phase transitions while Section~\ref{sec:krylov} introduces the theory of Krylov complexity of quantum states. The goal of these two sections is twofold: setting the notation of the paper and making it self-contained. Section~\ref{sec:lmg} offers an overview of the dynamical phase transition in the LMG model, and Section~\ref{sec:results} presents our results. We present our conclusions in Sec.~\ref{sec:conc}.

\section{Dynamical Quantum Phase Transition}
\label{sec:dynamical}

Let us consider a quantum many-body system described by a Hamiltonian $H(h)$, which depends on an externally controlled parameter $h$. At the initial time, the system is prepared in the ground state $\ket{\psi(0)}\equiv\ket{\psi_0}$ of the Hamiltonian $H(h_0) \equiv H_{0}$. After a sudden change, quench, during which the value of $h$ changes from $h_0$ to $h_f$, the system is left to evolve under $H(h_f) \equiv H_{f}$. In this context, the central quantity in the theory of dynamical quantum phase transition is the return probability amplitude
\begin{align}
    \mathcal{G}(t) = \overlap{\psi_0}{\psi_t} = \braket{\psi_0}{e^{-iH_ft}}{\psi_0},
\end{align}
also know as the survial amplitude \cite{Muga07,Muga12,delcampo11,delcampo16} and  the Loschmidt amplitude~\cite{Heyl2013}. 

The characterization of the DQPT relies on a formal analogy between $\mathcal{G}(t)$ and a special case of the boundary partition function $Z_B = \braket{\psi_1}{e^{-RH}}{\psi_2}$, with $R$ being the distance between the boundaries $\ket{\psi_1}$ and $\ket{\psi_2}$~\cite{Mussardo1995}.In this case, $R$ has to be interpreted as a spatiotemporal distance, in which one of its components is the complex time $it$ playing the role of the inverse temperature $\beta$, as guaranteed by relativistic quantum field theory. For more details, see also \cite{Andraschko2014}. Thereby, formally, $ \mathcal{G}(t)$ is the boundary partition function with $R=it$ and $\ket{\psi_1}=\ket{\psi_2}=\ket{\psi_0}$. The corresponding probability 
\begin{align}
    \mathcal{L}(t) = |\mathcal{G}(t)|^2.
\end{align}
is called return probability, survival probability, or Loschmidt echo.

The DQPT is then defined in terms of the Fisher or Lee-Yang zeros~\cite{Yang1952,Lee1952} of this partition function. Consider the return amplitude with complex time $t \to z = t + i\tau$. A transition occurs every time the zeros of $\mathcal{G}(z)$ crosses the real-time axis of the complex plane, and these crossings indicate the critical times of the dynamics. This analogy with the equilibrium quantum phase transition leads us to another important quantity in the dynamical case, a quantity that can be understood as the dynamical version of the free energy, the rate function
\begin{align}
    r(t) = -\frac{1}{N}\log\left[\mathcal{L}(t)\right],
\end{align}
with $N$ being the size of the system. At the critical times, $\mathcal{L}(t)$ is zero, and the rate function becomes nonanalytic. It is important to observe that this behavior only emerges in the thermodynamic limit where $N\rightarrow\infty$ and the Fisher zeros accumulate in a line or in a continuous region~\cite{HeylReview}.

The above-described phenomenon was termed DPT-II to differentiate it from another related phenomenon known as DPT-I. DPT-II was initially investigated in the transverse field Ising model~\cite{Heyl2013}, and since then, it has been explored in various quantum many-body systems. These encompass short- and long-range interacting quantum spin chains~\cite{Andraschko2014,Campbell2016,Zunkovic2016,Heyl2018,Hauke2021,Lang2018}, two-dimensional spin systems~\cite{Halimeh2022,Schmitt2015}, nonintegrable systems~\cite{Schuricht2013}, and optical systems~\cite{Puebla2020}, to name a few examples. For an in-depth review of DPT-II, covering both theoretical studies and experimental realizations, see Ref.~\cite{HeylReview}.

In the transverse field Ising model, the condition for the Fisher zeros to cross the real-time axis in the complex plane is that the quench crosses the critical point of the equilibrium quantum phase transition (QPT). Such a connection was thought to be a general feature of quantum many-body systems presenting a DQPT. However, several exceptions appeared later on, highlighting the nonexistence of a one-to-one correspondence between dynamical and equilibrium QPTs~\cite{HeylReview}. 

While DPT-II is investigated through the lens of the return probability, DPT-I is featured by the Landau order parameter of the system. This order parameter, a quantity whose derivative undergoes nonanalytic changes at a critical point of the quench parameter, serves to distinguish dynamical phases. Typically, the order parameter involves the time average of some physical quantity, such as the magnetization in spin chains. It is noteworthy that this dynamical critical point might not align with the critical point of the equilibrium QPT. DPT-I has been examined in various systems, and we point the reader to Refs.~\cite{Rey2022,Pappalardi2023} for comprehensive reviews on both its theoretical and experimental aspects.

Connections between DPT-I and DPT-II were also explored. As an example, in Ref.~\cite{Heyl2014}, a link is established between microscopic probabilities and the order parameter in the XXZ model when the initial state exhibits broken symmetry. In the transverse field Ising model with power-law decaying interaction, Ref.~\cite{Heyl2018} demonstrates that DPT-II occurs only when crossing the dynamical critical point of DPT-I. The study also reveals that the $\mathbb{Z}_2$ type symmetry, explicitly broken in the initial state, is restored in a long-time limit and at the critical times of DPT-II.

More recently, a connection between DPT-I and DPT-II and the excited-state quantum phase transition (ESQPT) in quantum many-body systems with infinite-range interactions has been established~\cite{Relano2022}. The authors in Ref.~\cite{Relano2022} define a generalized microcanonical ensemble by introducing three noncommuting charges and consider the presence of dynamical order parameters. In the thermodynamic limit, these order parameters are non-zero if the energy $E$ is less than the ESQPT critical energy $E_c$, but they always vanish when $E>E_c$. Concerning DPT-II, it was demonstrated that non-analyticities in the rate function $r(t)$ only occur if the energy of the system after the quench is greater than $E_c$, being absent if $E<E_c$. The same theoretical framework was applied to a finite-range interacting system in Ref.~\cite{Relano2023}.

The main focus of the present study is to investigate the complexity of the dynamics associated with DPT-I in the paradigmatic Lipkin-Meshkov-Glick (LMG) model~\cite{LMG1965}.

\section{Krylov Complexity of Quantum States}
\label{sec:krylov}

Next, we introduce a notion of complexity of the quantum dynamics based on the spreading of $\ket{\psi_t}$ over the system Hilbert space, or rather, the subspace spanned by the quantum evolution known as the Krylov space. For this reason, this quantity is also referred to as spread complexity~\cite{Wu2022,Caputa22}.

Let us start by writing the evolved state in the form
\begin{align}
    \ket{\psi_t} = \sum_{n=0}^\infty \frac{(it)^n}{n!} H^n\ket{\psi_0}.
\end{align}
The successive application of $H$ to the initial state generates the set of quantum states
\begin{align}
    \{\ket{\psi_0}, H\ket{\psi_0},H^2\ket{\psi_0},\cdots\} = \{H^n\ket{\psi_0}\}_{n\geq0},
\end{align}
which describes the spreading of the initial state over the Hilbert space during time evolution. The subspace spanned by $\{H^n\ket{\psi_0}\}_{n\geq0}$ is known as the Krylov space. The measure of complexity that we consider in this work was introduced in Ref.~\cite{Wu2022}, and it is given by
\begin{align}
    C_\mathcal{B}(t) = \sum_n c_n \vert\overlap{\psi_t}{B_n}\vert^2,
    \label{ComplexityB}
\end{align}
where $\mathcal{B}$ represents, for the time being,  a basis in the Hilbert space, whose elements are denoted as $\{\ket{B_n}\}_n$. Intuitively, we expect that complex dynamics lead to larger spreads over this basis. Thus, some constraints must be imposed on the coefficients $c_n$.  For a real and positive-definite measure of complexity, regardless of the state of the system, $c_n>0$ is required. The coefficients should vary with $n$ and grow monotonically with $n$ to assign higher complexity to components $\vert\overlap{\psi_t}{B_n}\vert^2$ with higher index $n$.
Note that the choice of constants coefficients $c_n=c$ is excluded as the complexity would reduce to the sum over all probabilities $p_n(t) = \vert\langle \psi_t \vert B_n \rangle\vert^2$ times $c$,  yielding $c$ at all $t$. 

A basis-independent complexity measure can be obtained from $C_\mathcal{B}(t)$ by performing a special minimization process, resulting in
\begin{align}
    C(t) = \min_{\mathcal{B}} C_\mathcal{B}(t).
    \label{Complexity}
\end{align}
Of course, it is always possible to construct a basis at the initial time such that $\ket{\psi_0} = \ket{B_0}$ has a non-zero overlap only with one state of the basis, thereby minimizing complexity. Instead, we consider a functional minimization of~\eqref{ComplexityB} which takes into account the spread of the state. For this purpose, it is natural to look at the set $\{H^n\ket{\psi_0}\}_{n\geq0}$ spanning the Krylov space since it includes only the portion of the Hilbert space visited by the system over time evolution.

The Krylov complexity, a precedent of the spread complexity, was first introduced in Ref.~\cite{Altman2019} in the context of operator evolution, i.e., in the Heisenberg picture. Based on this concept, a universal hypothesis for the maximum growth of local operators in quantum many-body systems was presented. Later, Balasubramanian and collaborators~\cite{Wu2022} extended the idea to the evolution of quantum states in the Schr\"odinger picture and proved that the basis which minimizes~\eqref{ComplexityB} for $c_n = n$ is the so-called Krylov basis. The Krylov basis is generated performing Gram-Schmidt orthogonalization on the set $\{H^n\ket{\psi_0}\}_{n\geq0}$. We denote the Krylov basis as $\mathcal{K}$ and its elements as $\{\ket{K_n}\}_n$.

Alternatively, the Krylov basis can be generated using the Lanczos algorithm~\cite{Lanczos1950,Muller2008}, which is a well-known recursive method used to generate an orthogonal basis. Starting with the initial state as the first Krylov state $\ket{K_0} = \ket{\psi_0}$, the next state is obtained as $\ket{K_1} = \frac{1}{b_1}H\ket{K_0}$ with $b_1 = \overlap{K_1}{K_1}^{1/2}$ being the normalization constant. The subsequent states $\{\ket{K_n}\}_{n\geq 2}$ are generated via the following recursion method
\begin{align}
    \label{Lanczos}
    &\ket{A_n} = H\ket{K_{n-1}} - a_{n-1}\ket{K_{n-1}} - b_{n-1}\ket{K_{n-2}},\\
    &\ket{K_{n}} = b_{n}^{-1}\ket{A_n}.
    \label{Eq:LanczosAlgorithm}
\end{align}
The constants $a_n$ and $b_n$, with $b_0=0$, are called Lanczos coefficients and they are defined as
\begin{align}
    a_n = \braket{K_n}{H}{K_n},\ \ \ \
    b_n = \overlap{K_n}{K_n}^{1/2}.
\end{align}
Isolating the first term in the right-hand side of Eq.~\eqref{Lanczos},
\begin{align}
    H\ket{K_{n-1}} = a_{n-1}\ket{K_{n-1}} + b_n\ket{K_n} + b_{n-1}\ket{K_{n-2}},
    \label{Eq:TridiagForm}
\end{align}
we observe that the Hamiltonian is tridiagonal in the Krylov basis $\{\ket{K_n}\}_n$. We note that the Lanczos algorithm can lead to roundoff errors \cite{Simon1984},  affecting the orthogonality between the states. For this reason, it may be necessary to re-orthogonalize the states.

The authors of Ref.~\cite{Wu2022} proved that the Krylov basis minimizes Eq.~\eqref{ComplexityB} in a specific way. Formally, let
\begin{align}
    S_\mathcal{B} = \left(C_\mathcal{B}^{(0)}, C_\mathcal{B}^{(1)},C_\mathcal{B}^{(2)},\cdots \right)
\end{align}
be the sequence of derivatives of $C_\mathcal{B}(t)$ calculated at $t=0$, that is
\begin{align}
    C^{(m)}_\mathcal{B} \equiv C_\mathcal{B}^{(m)}(0)  = \frac{d^m}{dt^m}C_\mathcal{B}(t)\Bigr|_{t=0},\ \ \ \ m=0,1,2,\cdots
\end{align}
We say that $S_{\mathcal{B}_1} < S_{\mathcal{B}_2}$ if there is some $k$ such that $C^{(m)}_{\mathcal{B}_1} = C^{(m)}_{\mathcal{B}_2}$ for $m<k$ and $C^{(m)}_{\mathcal{B}_1} < C^{(m)}_{\mathcal{B}_2}$ for $m=k$. Thus, for any basis $\mathcal{B}$, $S_{\mathcal{K}} \leq S_{\mathcal{B}}$ with the equality corresponding to the case $\mathcal{B} = \mathcal{K}$~\cite{Wu2022}. This is the minimization that we referred to above as functional minimization.

The Lanczos coefficients determine the matrix representation of the generator of evolution in Krylov space. Their specific role in quantum dynamics is still subject to investigation. The complexity growth in both pictures can be seen as a hopping single particle in a one-dimensional semi-infinite chain with the Lanczos coefficients representing the hopping terms. Thus, it is expected that complex dynamics makes the wavepacket of the hopping particle delocalize quickly in the semi-infinite chain. It is also for this reason that one chooses $c_n = n$ in the definition of complexity, since upon this choice, the complexity is the average position of the hopping particle in the semi-infinite chain. In what follows, we refer to this chain as the Krylov chain.


\section{DPT-I in the LMG model}
\label{sec:lmg}

Let us consider a spin chain described by the Hamiltonian
\begin{align}
    H(h) = -\frac{J}{N}S_z^2 - h S_x,
    \label{Eq:LMGHamiltonian}
\end{align}
where $S_\alpha = \sum_{i=1}^N \sigma^\alpha_i/2$ ($\alpha = x,y,z$) are collective spin operators, with $\sigma^\alpha_i$ denoting the $\alpha$ Pauli matrix acting on the $i$-th site of a chain of $N=2j$ sites and total angular momentum $j$. This Hamiltonian is a particular case of the well-known Lipkin-Meshkov-Glick model~\cite{LMG1965}. The constant $J$ denotes the ferromagnetic coupling between the spins in the $z$ direction while $h$ represents the strength of the magnetic field applied along the $x$-axis. This Hamiltonian describes an ensemble of $N$ spin-$1/2$ systems subject to all-to-all pairwise interactions in the presence of a magnetic field.

\begin{figure}[t!]
    \centering
\includegraphics[width=0.4\textwidth]{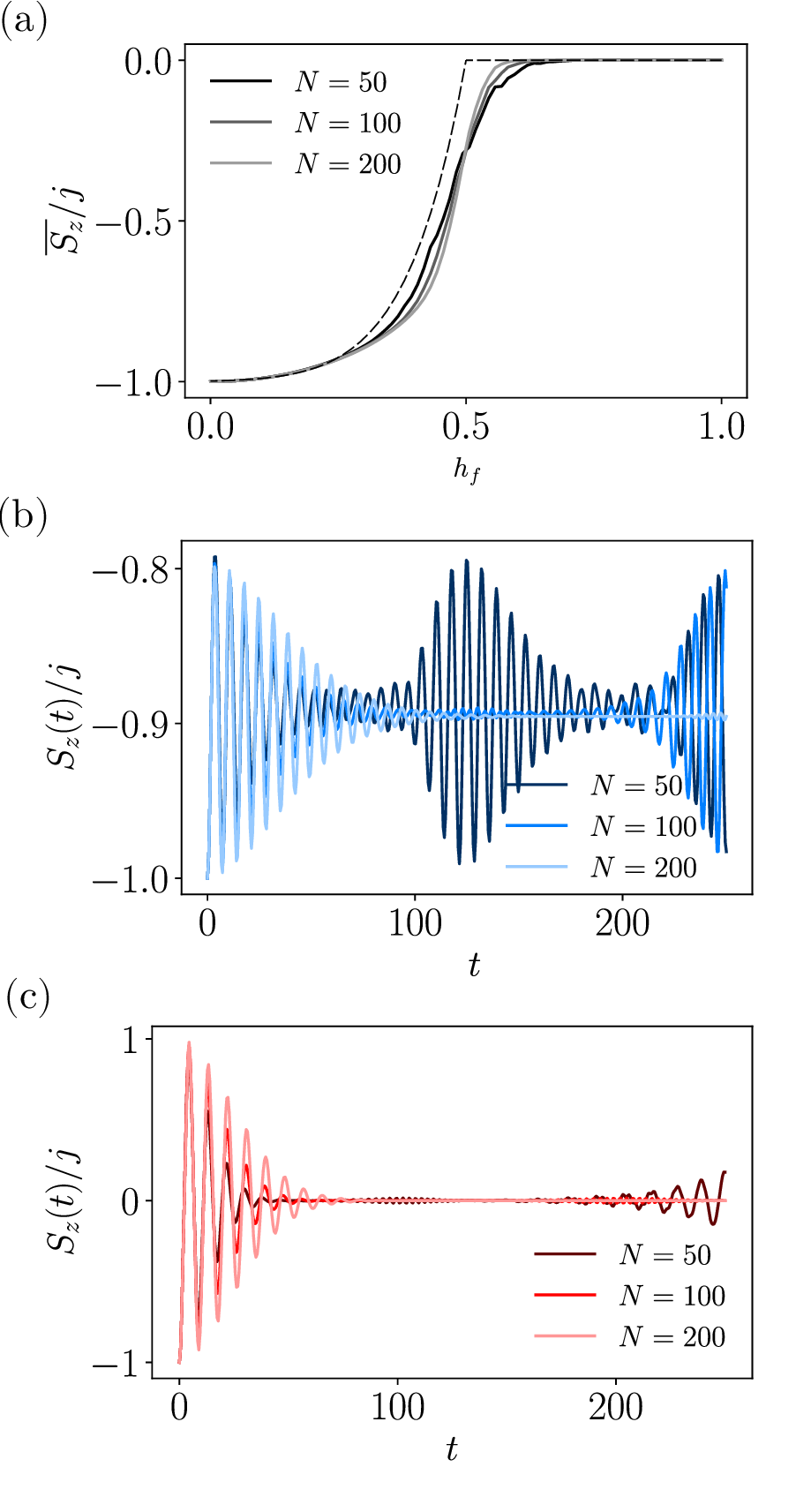}
    \caption{\textbf{Magnetization as an order parameter}. Panel (a) shows the time-averaged magnetization $\overline{S_z}$ as a function of quench $h_{0}=0 \rightarrow h_{f}=h$. The initial state is $\ket{\psi_0} = \ket{\downarrow}_z$. In the thermodynamic limit, $\overline{S_z}$ signals the dynamical phase transition at the critical point $h = 1/2$, thus identifying two dynamical critical phases. Panels (b) and (c) show the magnetization $S_z(t)$ as a function of time for a quench in the ferromagnetic phase ($h=0.3$) and a quench crossing the critical point to the paramagnetic phase ($h_f=0.8$), respectively. In both phases, revivals are suppressed in the thermodynamic limit.}
    \label{Fig1:OrderP}
\end{figure}

Considering a quench $h_{0} \rightarrow h_{f}>0$, we show in Fig.~\ref{Fig1:OrderP}(a) the bifurcation diagram of Hamiltonian~\eqref{Eq:LMGHamiltonian} for the time-averaged magnetization (order parameter for this model)
\begin{align}
    \overline{S_z} = \lim_{T\to \infty}\frac{1}{T} \int_0^T \braket{\psi(t)}{S_z}{\psi(t)} dt,
\end{align}
as a function of the quench parameter $h$, which characterizes DPT-I in this model. The initial state was taken to be the collective spin aligning with the $-z$ direction, the south pole of the Bloch sphere representation. This state corresponds to one of the two-fold degenerate ground states of $H_0$~\cite{Vidal2005}, denoted as $\ket{\downarrow}_z$. Throughout this paper, we use $J = 1$. The order parameter $\overline{S_z}$ delineates two distinct dynamical phases: a dynamical ferromagnetic phase for $h < 1/2$, where magnetization oscillates around a finite value depending on the initial state, leading to $\overline{S_z} \neq 0$, and a dynamical paramagnetic phase for $h \geq 1/2$, where magnetization oscillates around zero, yielding $\overline{S_z} = 0$. The critical point separating these phases is at $h_c = 1/2$.

The DPT-I in this model has been extensively studied~\cite{Heyl2018,Zunkovic2016,Silva2019}. The dynamical critical point, denoted as $h_c$, can be determined analytically for the Hamiltonian~\eqref{Eq:LMGHamiltonian}, resulting in $h_c = (h_0+J)/2$~\cite{Biroli2011}. As the system size increases, the behavior of $\overline{S_z}$ converges towards the mean-field solution, highlighting the exact nature of the mean-field solution for the LMG model in the thermodynamic limit~\cite{Vidal2005}. 

An essential aspect of DPT-I in this context is the symmetry of the initial state. This critical phenomenon is only observed when the initial state of the dynamics exhibits broken symmetry~\cite{Heyl2018}. The LMG model feature spin-flip symmetry~\cite{Ribeiro2008}, represented by the operator $\hat{\Pi} = e^{i\pi(\hat{S}_x-j)}$. This symmetry dictates that $H$ does not couple standard eigenbasis $\ket{j, m_z}$ states (eigenvectors of $S^2$ and $S_z$) with even and odd $m_z$. Under these conditions, a quench within the same dynamical phase produces oscillations in the system around a broken-symmetry effective state, while a quench crossing the dynamical critical point produces oscillations around a symmetric effective state. These oscillations are characterized by long-lived steady states, and they are usually linked to prethermalization~\cite{Ueda2018}. 

Figures~\ref{Fig1:OrderP}(b) and~\ref{Fig1:OrderP}(c) illustrate instances of magnetization behavior over time, depicting a quench within the same phase and crossing the dynamical critical point, respectively. The observed revivals in these figures arise from finite-size effects within the system.  As the system size $N$ increases, these revivals disappear.
In the following section, we demonstrate that the time-averaged Krylov complexity can also serve as an order parameter for this system.

\section{Results}
\label{sec:results}

We start by discussing the spread complexity's behavior and its long-time average. To understand why Krylov complexity can be taken as an order parameter for the DPT-I in the LMG model, we proceed with the analysis of two other quantities, the inverse participation ratio, and the Shannon entropy, considering the energy and the Krylov basis. 

\subsection{Krylov Complexity of quantum states in the LMG model}

\begin{figure}
    \centering
    \includegraphics[width=0.45\textwidth]{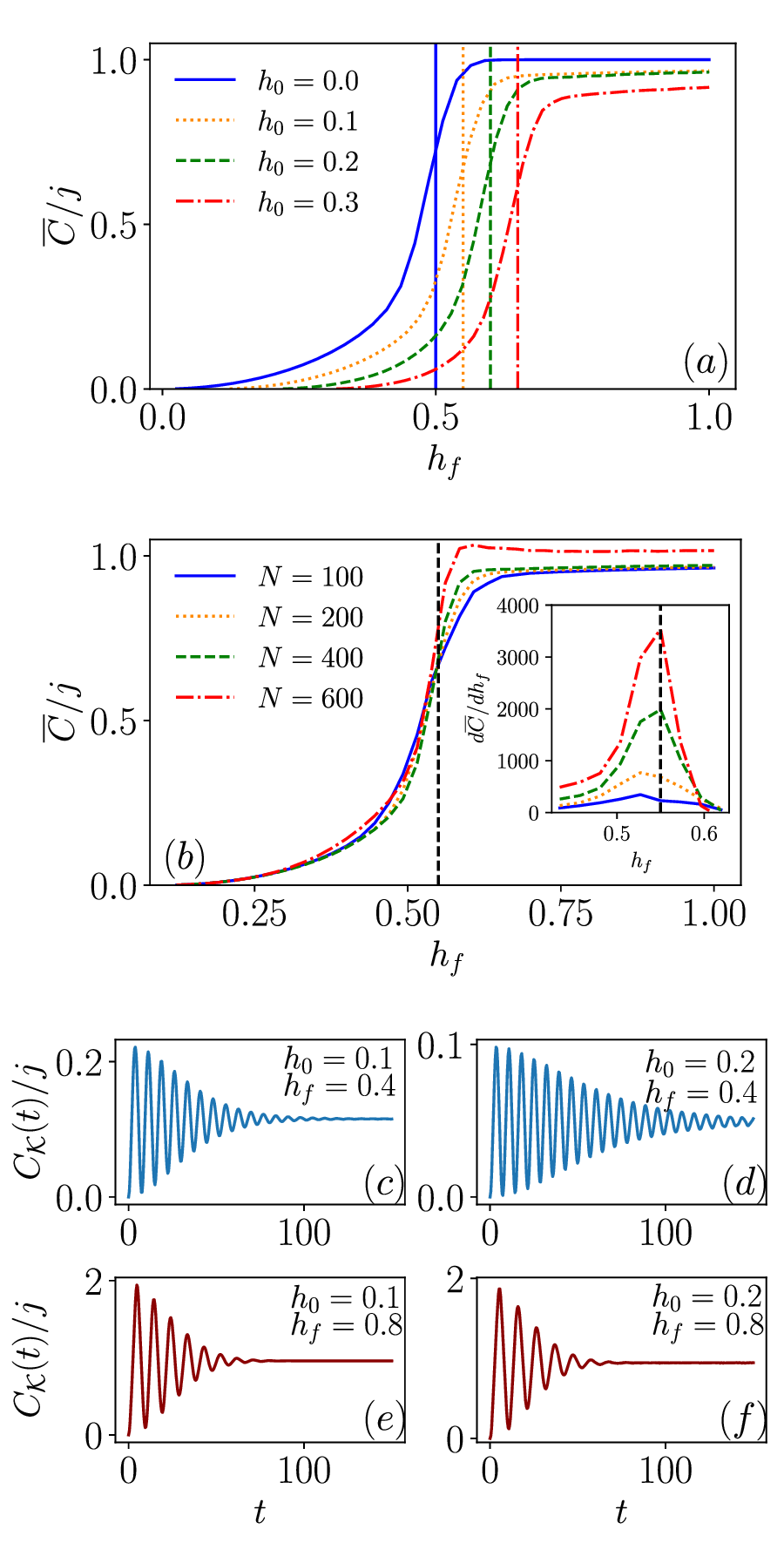}
    \caption{\textbf{Krylov complexity as an order parameter}. Panel (a) shows the normalized time-averaged Krylov complexity $\overline{C}/j$ as a function of $h_f$ for several values of $h_0$ and $N=200$. $\overline{C}$ exhibits exactly the same transition as $\overline{S_z}$ at the dynamical critical point given by $h_c = (h_0+J)/2$ and represented by the vertical lines. Panel (b) shows the normalized time-averaged Krylov complexity $\overline{C}/j$ for $h_0 = 0.1$ and several values of $N$ as a function of $h_f$. As $N$ increases, the non-analytical point of $\overline{C}$ approaches the dynamical critical point $h_c = (h_0+J)/2 = 0.55$ (black vertical line), indicating that $\overline{C}$ exhibits the expected behavior of an order parameter in the thermodynamic limit ($N\to \infty$), while the inset shows its normalized derivative w.r.t. $h_f$ reaching higher values at $h_c$ as $N$ increases. Panels from (c) to (f) show examples of $C_\mathcal{K}(t)$ as a function of time for different quenches within the same phase ((c),(d)) and crossing $h_c$ ((e),(f)). We see that $C_\mathcal{K}(t)$ exhibits the same pattern of oscillations as $S_z(t)$. In all panels, we used $\ket{\psi_0} = \ket{\downarrow}_z$, and in panels (a) and (b), we used $T=150$.}
    \label{Fig2:TimeAve_Complexity}
\end{figure}

We turn to the spread complexity given in Eq.~\eqref{ComplexityB}. We argue that the long-time average of the Krylov complexity
\begin{align}
    \overline{C} = \lim_{T\to \infty} \frac{1}{T} \int_0^T C_\mathcal{K}(t)dt
    \label{Eq:TimeAveragedC}
\end{align}
behaves as a dynamical order parameter in the DPT-I of this model. 

Using the setup described in the last section,   Fig.~\ref{Fig2:TimeAve_Complexity}(a) shows the normalized time-averaged complexity as a function of the quenching intensity $h_0<h_f\leq1$ for several values of $h_0$. Remarkably, $\overline{C}$ exhibits the same qualitative behavior as the dynamical order parameter $\overline{S_z}$, with the only difference being the numerical values assumed in both dynamical phases. 

Moreover, we note that $\overline{C}/j$ exhibits the same dynamical critical point as $\overline{S_z}$, whose dependence with $h_0$ is $h_c = (h_0+J)/2$. This is ensured by the fact that as $N$ increases, the non-analytical point of $\overline{C}$ approaches the exact value $h_c$, as can be seen from Fig. \ref{Fig2:TimeAve_Complexity}(b) for an example with $h_0=0.1$, which implies that as $N$ increases the derivative of $\overline{C}$  with respect to $h_f$ reaches even higher values and shall diverge in the thermodynamic limit $N\to \infty$ (see the inset at Fig. \ref{Fig2:TimeAve_Complexity}(b)). This fact was also verified for other values of $h_0$ and holds true.

As with the magnetization $S_z(t)$, the Krylov complexity exhibits distinct behavior in each dynamical phase over time. This characteristic is depicted in the panels (c)-(f) of Fig.~\ref{Fig2:TimeAve_Complexity}. Extensive numerical calculations suggest that, in the thermodynamic limit, and for $h_0=0$, the complexity oscillates around a finite value, consistently remaining below unity for a quench within the ferromagnetic phase. Conversely, it oscillates and later stagnates on the unit value for a quench into the paramagnetic phase. As we increase $h_0$, we observe the same pattern for quenches within the same phase, however it stagnates on progressively smaller values than unity, hence making its time-average, $\overline{C}$, display smaller plateaus as can be seen from Fig. \ref{Fig2:TimeAve_Complexity}(a) for $h_f>h_c$. Despite a smaller plateau, $\overline{C}$ seems to reach a constant and maximum value for $h_f > h_c$, which suggests that DPT-I causes the maximum spreading regime of the dynamics. Based on these findings, we can argue that we use the time-averaged Krylov complexity to indicate a DPT-I.

We additionally examined how the time-averaged complexity varies concerning the ground-state manifold. The LMG model possesses a two-fold degenerate ground state, representing the north and south poles in the Bloch sphere representation of the collective spin variables $S_\alpha,\ (\alpha = x, y, z)$. Interestingly, both of these initial ground states lead to the same pattern in the behavior of the Krylov complexity over time. Consequently, the long-time average $\overline{C}$ as a function of $h$ exhibits a qualitative similarity to $\overline{S_z}$.

Given that DPT-I is associated with a dynamical symmetry breaking, it is reasonable to anticipate that the sensitivity of the Krylov basis extends to the symmetry of the model. Therefore, a deeper relationship between the Krylov basis and the energy one should exist. This aspect will be investigated in the concluding part of this section.

\begin{figure}
    \centering
    \includegraphics[width=0.39\textwidth]{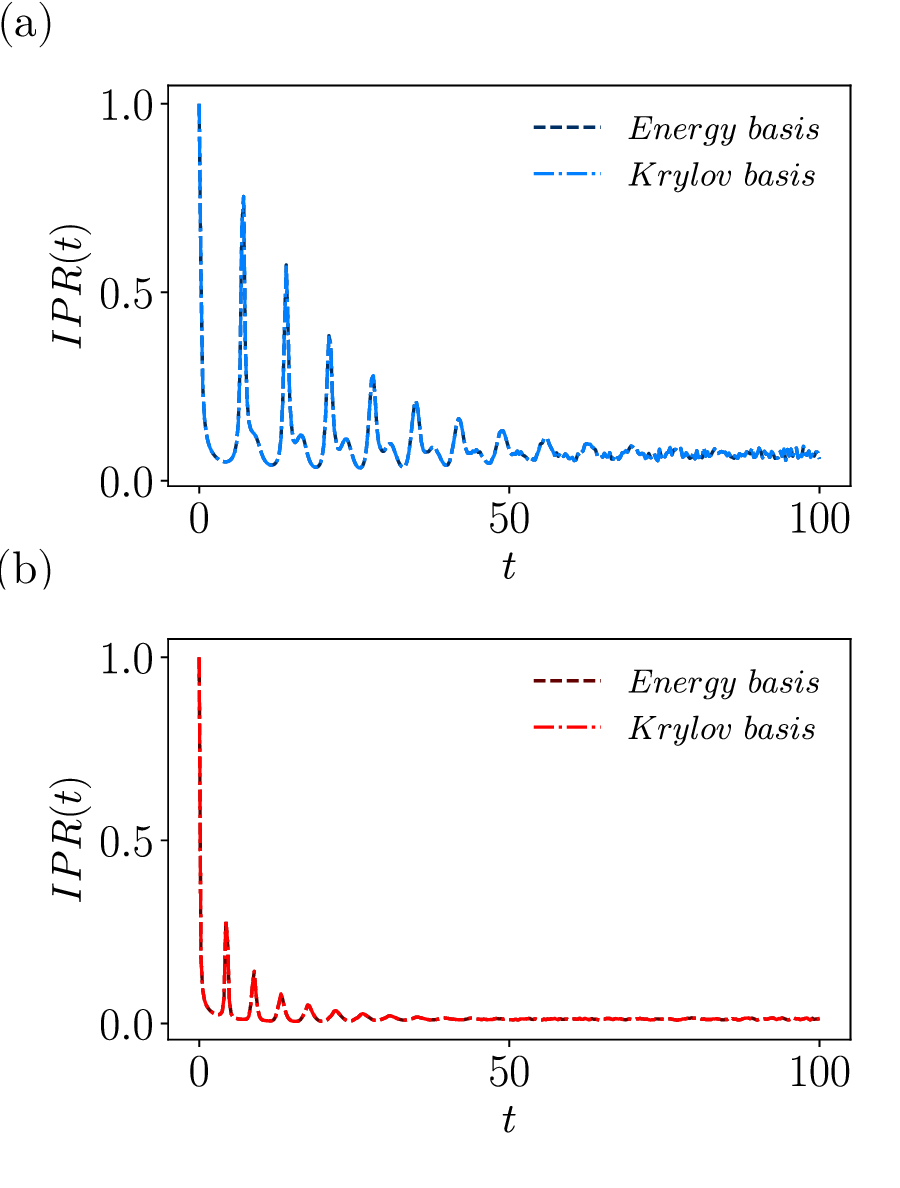}
    \caption{\textbf{Inverse Participation Ratio}. Panel (a) shows IPR for a quench $h_f = 0.3$, while panel (b) shows the same quantity but for a quench crossing the critical point ($h_f=0.8$). We take $N=200$ in both panels. Exactly the same behavior is observed if the initial state is taken as the second ground-state, $\ket{\uparrow}_z$.}
    \label{Fig:IPR}
\end{figure}

\subsection{Inverse Participation Ratio}

To analyze the relation between the energy basis and the Krylov basis, we first consider the inverse participation ratio (IPR) in each one of these bases. The IPR is given by
\begin{align}
    \text{IPR}(t) = \sum_k \vert\overlap{k}{\psi(t)}\vert^4 = \sum_k p_k^2(t),
    \end{align}
for some basis whose elements we generically denote as $\{\ket{k}\}$. As its name suggests, the IPR measures how many states of the chosen basis effectively participate in the course of the time evolution of the system. Using the analogy with the Krylov chain, it means that the IPR should measure the level of localization of $\ket{\psi_t}$ in this chain for a non-trivial time evolution since at $t=0$, $\ket{\psi_0}$ is sharply localized in the first site of the Krylov chain as demands the first step of the Lanczos algorithm.

Considering the same protocol as before, we compute the IPR for two distinct bases: the pre-quench energy eigenbasis (the eigenvectors of $H_0$) and the Krylov basis. Two instances of the results are shown in Fig.~\ref{Fig:IPR}. Interestingly, the IPR computed in both bases are identical. However, this coincidence only happens when we start with $h_0 = 0$, when the initial energy spectra is doubly degenerate. Extensive numerical analysis shows that they are different otherwise, see appendix $\ref{Appendix:IPR_Shannon}$ for some examples of this mismatch. Another property of IPR that we observed numerically is the independence with respect to which ground-state is taken as the initial state, thus giving the same result also for the second ground-state $\ket{\psi_0} = \ket{\uparrow}_z$.

Other two aspects of the IPR are noteworthy. The first one is the appearance of periodic peaks that decay in amplitude along the dynamics. These peaks are due to partial revivals in the dynamics, that is, the state of the system periodically returns almost completely to the initial state, a feature expected for systems that exhibit DQPT including the LMG model \cite{lerma2018,Campbell2016}. This fact can also be seen by noticing that the first term of the sum in the IPR is the squared survival probability
\begin{align}
    \text{IPR}(t) = \mathcal{L}^2(t) + \sum_{k>0}p_k^2(t),
\end{align}
which characterizes DQPT as described in the section \ref{sec:dynamical}, and the relation above holds for both the pre-quench energy basis and the Krylov basis in our protocol. Therefore, the IPR is also controlled by the survival probability. The second aspect of the IPR that we highlight is the decay rate. We observe that the closer the quench from the dynamical critical point, the more states participate in the dynamics, and hence, the faster the IPR decreases. An increase in the number of Krylov states effectively participating in the dynamics was observed for quenches crossing the equilibrium QPT of the LMG model in the Ref. \cite{afrasiar2023} as well.

The reason for the match when $h_0=0$ will be analytically explored later in this article. However, before presenting this analysis, let us consider another interesting quantity, the Shannon entropy.

\subsection{\texorpdfstring{$\mathcal{K}$}{TEXT}-Entropy - Shannon entropy in the Krylov basis}

Considering a probability distribution $p_{n}$, the Shannon entropy
\begin{align}
    \mathcal{E}(t) = -\sum_n p_n(t)\log(p_n(t))
\end{align}
is a measure of the uncertainty of the system or the classical information contained in the system. We are here interested in the entropy associated with both bases discussed in the previous section: the Krylov basis, for which $p_n(t) = |\overlap{K_n}{\psi(t)}|^2$ and the initial energy basis, whose probabilities are $p_n(t) = |\overlap{E^0_n}{\psi(t)}|^2$. 
The Shannon entropy in the Krylov basis was introduced in Ref.~\cite{Barbon19} and is also known as the $\mathcal{K}$-entropy. It is a measure of the complexity of the dynamics. The authors of Ref.~\cite{Wu2022} argued that the complexity defined as the exponential of this quantity measures the minimum Hilbert space dimension required to store the probability distribution $p_n(t)$. 

\begin{figure}[t!]
    \centering
    \includegraphics[width=0.4\textwidth]{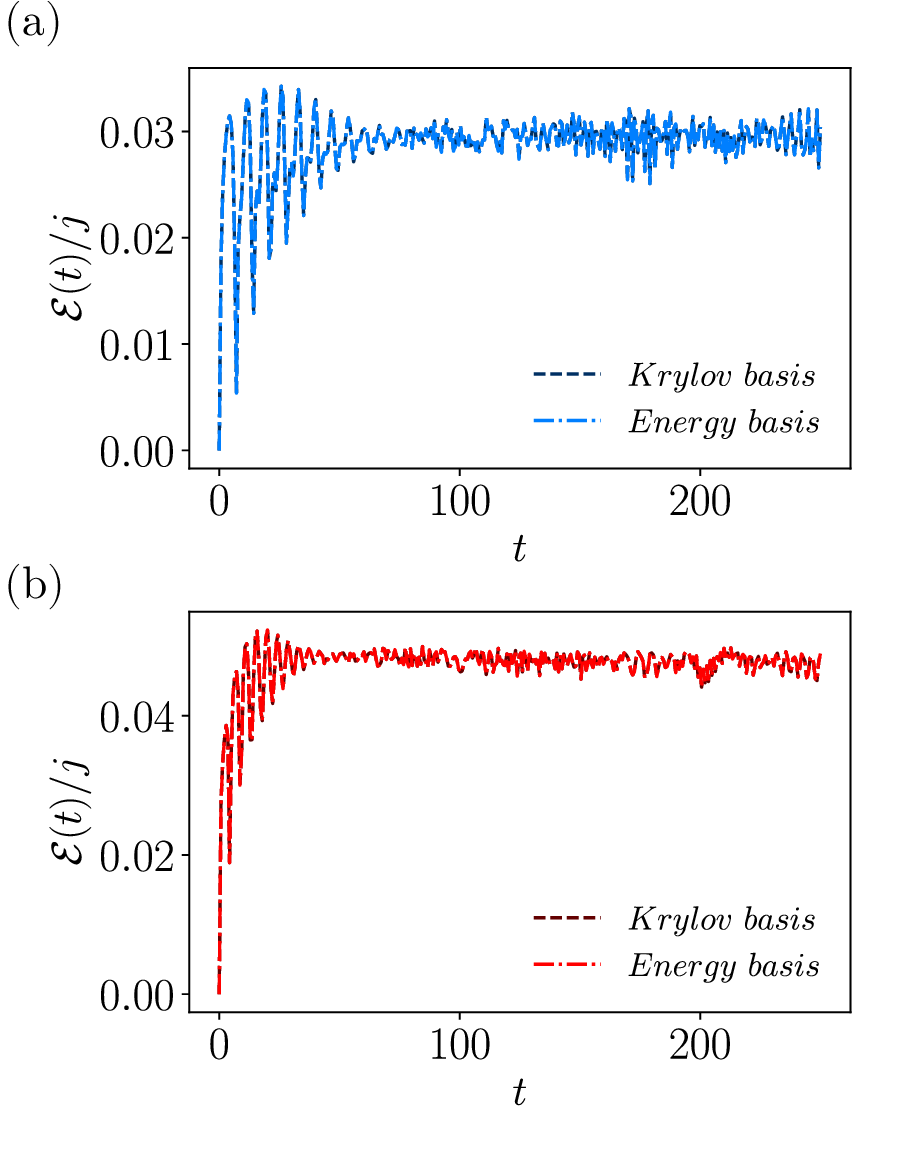}
    \caption{\textbf{Shannon entropy}. Panel (a) shows the results for the quench $h = 0.3$, while panel (b) considers the case $h = 0.8$. In both cases, the initial state is $\ket{\psi_0}=\ket{\downarrow}_z$ and $N=200$ as in the previous calculations. Similar results are obtained considering the second ground-state of the LMG model for $h_0=0$, which we denote by $\ket{\uparrow}$.}
    \label{Fig4:ShannonEntropy}
\end{figure}

By comparing the Shannon entropy in both bases, a perfect match is again observed. Also, such a feature only occurs when we start at $h_0 = 0$. An example is shown in Fig.~\ref{Fig4:ShannonEntropy} for the same parameters used in Fig.~\ref{Fig:IPR} and we show instances of deviation from this behavior when $h_0\neq 0$ in the appendix \ref{Appendix:IPR_Shannon}.

We note that the probabilities $p_n(t) = |\overlap{E^0_n}{\psi(t)}|^2$ are the populations of the density matrix in the energy eigenbasis and, thus, the entropy associated with such distribution is the diagonal entropy, proposed as the thermodynamic entropy for closed quantum systems~\cite{Polkovnikov2011}. Since $p_n(t) = |\overlap{K_n}{\psi(t)}|^2$ can be seen as the probability density associated with site $n$ in the semi-infinite Krylov lattice, naturally, the corresponding Shannon entropy $\mathcal{E}_\mathcal{K}(t)$ can be interpreted as the uncertainty in the spreading of the initial state through the Krylov subspace.

The numerical results described above indicate a deeper match beyond the average spread. The goal of the next subsection is to investigate why this happens. 

\subsection{Derivation of the Krylov basis for \texorpdfstring{$h_0=0$}{TEXT}}

Let us consider quenches starting from $h_0 = 0$. In this case, it is possible to derive analytical results involving the Krylov basis. The eigenstates of the pre-quench Hamiltonian
\begin{align}
    H_0 = -\frac{1}{2j}S_z^2,
\end{align}
are the usual angular momentum basis $\ket{j,m_z}$, with the index $m_z \in \{-j,-j+1,\cdots,j-1,j\}$ specifying the $2j+1$ spin projections. Of course, the ground-states denoted by $\ket{\uparrow}_z$ and $\ket{\downarrow}_z$ mentioned previously are examples of these eigenstates. Then, the quench is performed in the magnetic field $h$ such that the post-quench Hamiltonian can be written in terms of ladder operators
\begin{align}
    H_f = -\frac{1}{2j}S_z^2 - \frac{h_f}{2}\left(S_+ + S_-\right).
    \label{Eq:PosQuenchH}
\end{align}

Now we observe that the set of operators $\{S_z,S_+,S_-\}$ are the operators defining the well-known $SU(2)$ algebra, thus the following commutation relations hold
\begin{align}
    [S_z,S_{\pm}] = \pm S_{\pm},\ \ \ \ [S_+,S_-] = 2S_z.
\end{align}

The action of the Hamiltonian given in Eq.~\eqref{Eq:PosQuenchH} on one of the basis states $\ket{j,m_z}$ is
\begin{align}
    H_f\ket{j, m_z} = c_0\ket{j,m_z} + c_{+}\ket{j,m_z+1} + c_{-} \ket{j,m_z-1},
    \label{Eq:HfAction}
\end{align}
where $c_0 = -m_z^2/2j$, $c_+ = -\frac{h}{2}\sqrt{j(j+1) - m_z(m_z+1)}$ and $c_{-} = -\frac{h}{2}\sqrt{j(j+1) - m_z(m_z-1)}$. Comparing Eqs.~\eqref{Eq:HfAction} with Eq.~\eqref{Eq:TridiagForm}, we immediately see that the states satisfying the Lanczos algorithm for $\ket{\psi_0} = \ket{j,m_z}$ are precisely the set $\{\ket{j,m_z}\}$ up to a factor $\pm1$, that is
\begin{align}
    \ket{K_i} = \pm\ket{j,m_z}.
    \label{Eq:KrylovBasis}
\end{align}
Therefore, we conclude that the Krylov states are proportional to the pre-quench energy eigenstates. This explains why both the IPR and the Shannon entropy are the same in both bases.

\begin{figure}
    \centering
    \includegraphics[width=0.45\textwidth]{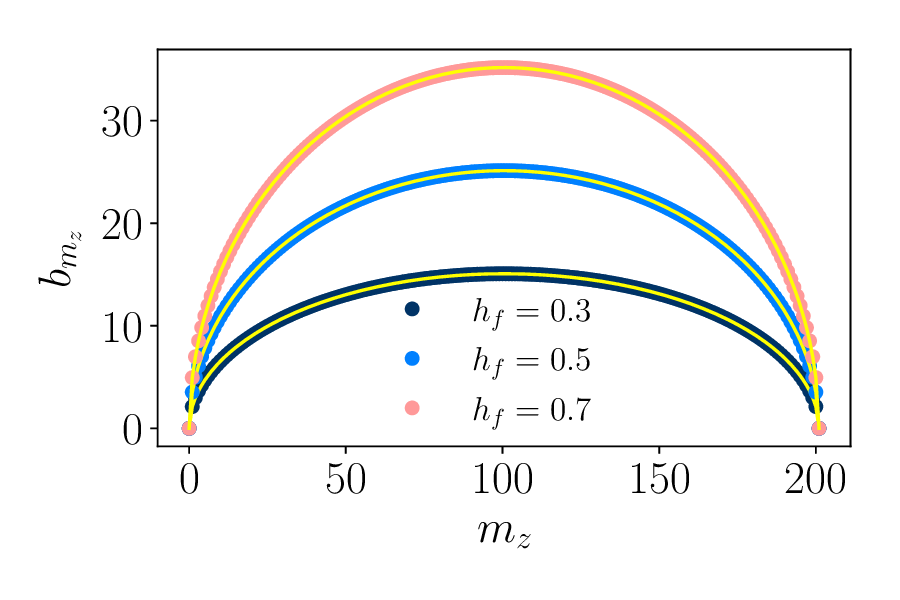}
    \caption{\textbf{Lanczos coefficients}. Comparison between the numerical (dots) and analytical (yellow lines) coefficients for $N=200$ and $\ket{\psi_0}=\ket{\downarrow}_z$. They feature perfect agreement showing that $b_{m_z} = c_{+}(m_z)$ and attesting that the Krylov states are the usual angular momentum states $\ket{j,m_z}$ if the initial state is equal to any one of the ground-states.}
    \label{Fig:Lanczos}
\end{figure}

The action of $H_f$ on the state $\ket{j,m_z}$, Eq.~\eqref{Eq:HfAction}, provides us an analytical expression for the Lanczos coefficients, which correspond to the constant $c_{-}$. For convenience, we relabel the index $m_z$ as $m_z \rightarrow -j + m_z$ such that now $m_z$ runs through the set $\{0,1,\cdots,2j\}$. With this change, the constant $c_{-}$ becomes
\begin{align}
    c_{-}(m_z) = \frac{h_f}{2}\sqrt{m_z(2j-m_z+1)},
    \label{Eq:LanczosAnalytic}
\end{align}
where we absorbed the minus sign in the state $\ket{j,m_z}$. Figure~\ref{Fig:Lanczos} shows the perfect agreement between the Lanczsos coefficients calculated numerically using the Hamiltonian~\eqref{Eq:LMGHamiltonian} and the Lanczos algorithm~\eqref{Eq:LanczosAlgorithm} and the analytical expression for $c_{-}$, proving that $c_{-}(m_z) = b_{m_z}$.

Since $\ket{K_{m_z}} = \pm\ket{j,m_z}$, the Krylov basis must share the same symmetries as the pre-quench energy basis. This fact turns the Krylov basis sensitive to the break or restoration of the spin-flip symmetry if a quench is performed in the LMG model, which explains the ability of the Krylov complexity to characterize DPT-I. Moreover, note that we can write the expression of the Krylov complexity in the form
\begin{align}
    C_\mathcal{K}(t) = \sum_{m_z=0}^{2j} m_z \vert\overlap{\psi_t}{K_{m_z}}\vert^2.
\end{align}
Considering now the expression of $S_z(t)$,
\begin{align}
    S_z(t) = \braket{\psi_t}{\hat{S}_z}{\psi_t}.
\end{align}
and employing the completeness of the angular momentum basis, $\sum_{m_z=0}^{2j}\ket{j,-j+m_z}\bra{j,-j+m_z} = \mathbb{I}$, we can readily show that
\begin{align}
    C_\mathcal{K}(t) = S_z(t) + j,
\end{align}
thus confirming that
\begin{align}
    \overline{C} = \overline{S_z} + j.
\end{align}

The above discussion proves that the Krylov complexity must have the same time behavior as the magnetization. We thus conclude that its time average is an order parameter for this model. 

Finally, we note that when $\ket{\psi_0} = \ket{j,m_z}$ for any $m_z$, $\dim{(\mathcal{K})}=\dim{(\mathcal{H})}$, i.e., the dimension of the Krylov subspace is equal to the dimension of the Hilbert space. These initial states explore the whole Hilbert space, showing a kind of ergodicity. We emphasize that this is not a feature of all quantum systems. 

\section{Conclusion}
\label{sec:conc}

Understanding the complex nature of the temporal evolution of many-body quantum systems holds fundamental significance across various research fields. This study contributes to this area by examining the Krylov complexity (spread complexity) and its connection to the dynamical phase transition within the LMG model. In essence, we show that the time-averaged Krylov complexity acts as an order parameter in this context. This conclusion stems from a numerical investigation encompassing not only the Krylov complexity but also the inverse participation ratio and Shannon entropy within the Krylov basis. This analysis suggests a relation between the Krylov basis and the energy eigenbasis. A further analytical study establishes that the time-averaged Krylov complexity effectively signals the dynamical phase transition in this model using the equivalence between the Krylov and the energy bases. 

It is important to note that a thorough numerical investigation reveals that this connection is valid exclusively in instances where the symmetry of the model is broken, as shown analytcially. This prompts questions about how the Krylov complexity behaves under general changes of symmetry. Addressing this question will certainly deepen our comprehension of the dynamics inherent in many-body systems.

Our findings also provide insights into the thermodynamics of critical systems of this nature. While the thermodynamics of equilibrium phase transitions is well-established, the same cannot be said for its dynamic counterpart. Our results indicate that, in the case of the LMG model, the $\mathcal{K}$-entropy serves as the quantum thermodynamic entropy. This assertion is grounded in the definition of diagonal entropy \cite{Polkovnikov2011} and the gauge theory of quantum thermodynamics outlined in Ref. \cite{Celeri2021}. Furthermore, this observation aligns with earlier investigations into the thermodynamics of DQPTs, where a thermodynamic entropy in quantum phase space not only signals the transition but also, on average, exhibits a monotonic increase over time~\cite{Goes2020}. Consequently, our results suggest a profound connection between DPQTs and the process of thermalization.

An intriguing observation emerges when considering the interconnectedness of symmetry and thermalization in thermodynamics, suggesting a profound link between the inquiries addressed in the preceding paragraphs.

Whether the time-averaged Krylov complexity can function as an order parameter for DPT-I in systems beyond the LMG model remains an open question. Additionally, elucidating the broader applicability of the connection between Krylov and energy bases is crucial. Addressing these questions should provide valuable insights into the underlying nature of this dynamical critical behavior.

\begin{acknowledgments} 
The authors are thankful to Anatoly Dymarsky and Pratik Nandy for their feedback on the manuscript. This work was supported by the National Institute for the Science and Technology of Quantum Information (INCT-IQ), Grant No. 465469/2014-0, the National Council for Scientific and Technological Development (CNPq), Grants No. 308065/2022-0, and the Coordination of Superior Level Staff Improvement (CAPES).
\end{acknowledgments}


\appendix
\section{IPR and \texorpdfstring{$\mathcal{K}$}{TEXT}-entropy in the case \texorpdfstring{$h_0 \neq 0$}{TEXT}}
\label{Appendix:IPR_Shannon}

\begin{figure}[t]
    \centering
    \includegraphics[width=0.5\textwidth]{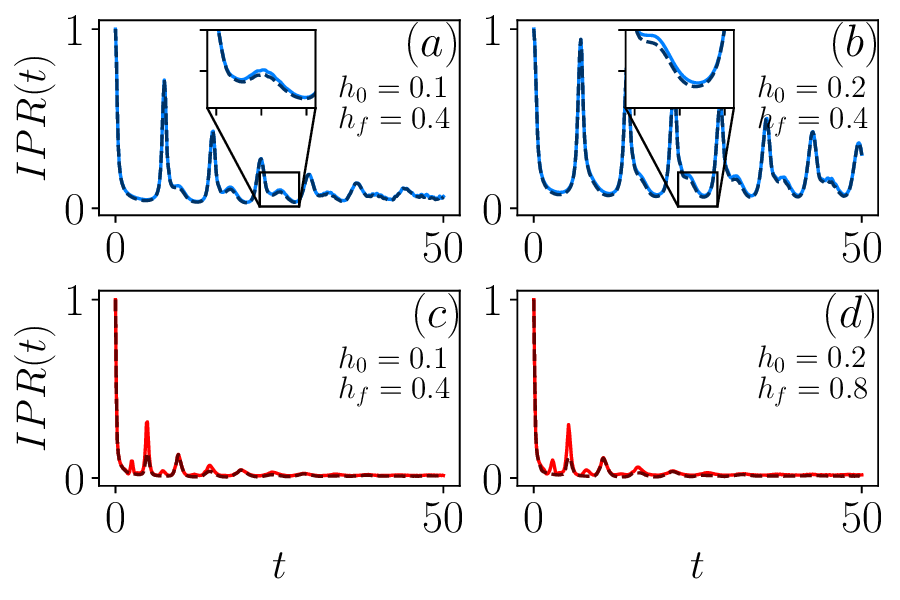}
    \caption{\textbf{Inverse participation ratio when $h_0\neq 0$}. Panels (a) and (b) show IPR for quenches lying in the dynamical ferromagnetic phase, while panels (c) and (d) show IPR for quenches crossing the dynamical critical point for $h_0\neq 0$. The inset in panel (a) only shows a zoom of an interval of the curves to help in the visualization. In all the panels, the solid line refers to the energy basis, while the dashed line refers to the Krylov basis ($N=200$). We observe a mismatch between the curves for all the cases, which indicates that the pre-quench energy basis and the Krylov basis are different from each other when $h_0\neq 0$.}
    \label{Fig6:IPR_h0!=0}
\end{figure}
\begin{figure}[t]
    \centering
    \includegraphics[width=0.5\textwidth]{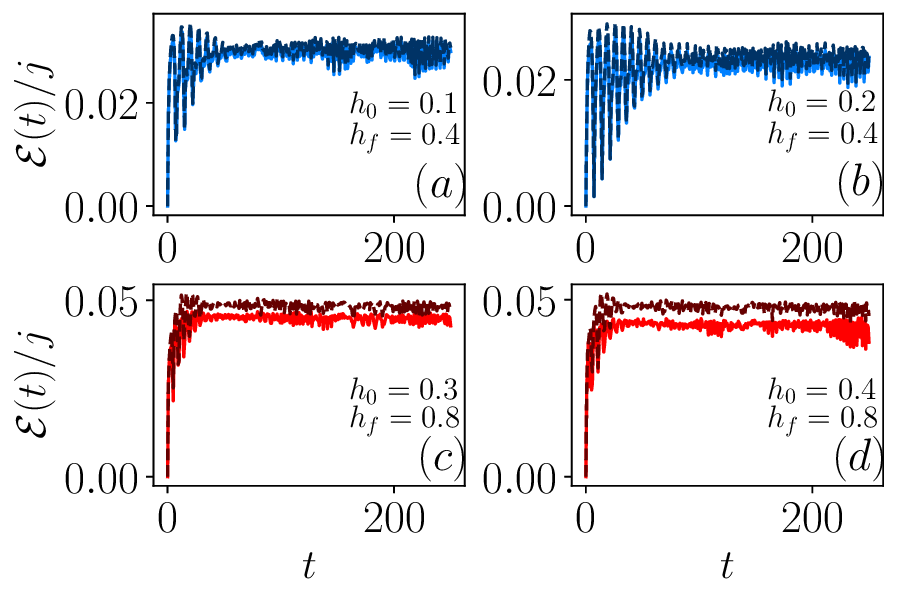}
    \caption{\textbf{Shannon entropy when $h_0\neq 0$.} Panels (a) and (b) show the Shannon entropy for quenches lying in the dynamical ferromagnetic phase and panels (c) and (d) show the Shannon entropy for quenches crossing the dynamical critical point ($N=200$). In all the panels, the solid line refers to the pre-quench energy basis, while the dashed line refers to the Krylov basis. Similarly to the IPR, we can see a clear mismatch between the Shannon entropy calculated in both bases when $h_0\neq 0$.}
    \label{Fig:ShannonE_h0!=0}
\end{figure}

In this appendix, we discuss the IPR and the Shannon entropy in quenches starting at $h_0 \neq 0$. Figure \ref{Fig6:IPR_h0!=0} shows four different instances of quenches starting at $h_0 \neq 0$ and the respective evolution of the IPR. In all these cases, a mismatch is observed between the IPR in the pre-quench energy basis (solid lines) and in the Krylov basis (dashed lines). This mismatch happens irrespective of whether the quench lies in the same phase as the pre-quench hamiltonian or if the quench crosses the dynamic critical point. Similar features can be seen in the Shannon entropy. For the same values of $h_0$ and $h_f$ as in Fig. \ref{Fig6:IPR_h0!=0}, figure \ref{Fig:ShannonE_h0!=0} displays the mismatch between the Shannon entropy in the pre-quench energy basis (solid lines) and in the Krylov basis (dashed lines).

Interestingly, as  Fig. \ref{Fig6:IPR_h0!=0} indicates, $\ket{\psi_t}$ displays greater delocalization in the Krylov basis than in the energy basis when $h_0 \neq 0$. This result is in accordance with the respective behavior of the Shannon entropy since greater delocalization in the Krylov basis shall imply greater uncertainty regarding the spread of $\ket{\psi_0}$ through the Krylov subspace, exactly what we see from Fig. \ref{Fig:ShannonE_h0!=0}.

\bibliography{KC_LMG}

\end{document}